\documentclass[12pt,russian]{article}

\usepackage[T2A]{fontenc}
\usepackage{amsmath,amsfonts,amssymb,amsthm}
\usepackage[russian,english]{babel}

\newcommand{\bgeq}{\begin{equation}}
\newcommand{\edeq}{\end{equation}}

\newcommand{\R}{\mbox{\bf R}}
\renewcommand{\C}{\mbox{\bf C}}

\newcommand{\N}{\mbox{\bf N}}

\newcommand{\prf}{{\par\noindent\bf Доказательство. }}

\newcommand{\edprf}{\hspace{5em}\rule{.5em}{.5em}}
\newcommand{\edlm}{\par\noindent Лемма доказана. \edprf}

\newcommand{\edth}{\par\noindent Теорема доказана. \edprf}

\newcommand{\re}{\mathop{\rm Re}}
\newcommand{\im}{\mathop{\rm Im}}

\newcommand{\sgn}{\mathop{\rm sgn}\nolimits}

\newcommand{\kan}{\thicklines}
\newcommand{\nan}{\thinlines}
\newtheorem{thrma}{Теорема}[section]
\newtheorem{lm}{Лемма}[section]
\newtheorem{cl}{Следствие}[section]
\newtheorem{df}{Определение}[section]
\newtheorem{nt}{Замечание}[section]
\newtheorem{pr}{Предложение}[section]
\newcommand{\sqplus}{\sqrt{\lambda}}
\newcommand{\la}{\lambda}

\newcommand{\sa}{{\Sigma}}

\newcommand{\tttttt}{\gamma}
\newcommand{\tapich}{\gamma_0}
\newcommand{\taa}{\gamma_a}
\newcommand{\taam}{\gamma_a^-}
\newcommand{\tab}{\gamma_b}
\newcommand{\tainf}{\gamma_\infty}

\renewcommand{\No}{N}

\begin{document}

\begin{center}

{\LARGE\bf
On the Model for the Orr--Sommerfeld Equation \\[0.3cm]
with Quadratic Profile
}

\vspace{10mm}

{\large
A. A. Shkalikov$^{(a)}$ and S. N. Tumanov$^{(b)}$
}

\vspace{10mm}

Department of Mechanics and Mathematics,\\
Moscow State University,\\
Moscow, Russia\\
{\em
${}^{(a)}$e-mail: ashkalikov@yahoo.com\\
${}^{(b)}$e-mail: tumanov@art-com.ru\\
}

\vspace{15mm}

{\bf Abstract}

\end{center}

\noindent
The model for Orr--Sommerfeld equation with quadratic profile on the finite interval
is considered. The behavior of the spectrum of this problem is completely
investigated for large Reynolds numbers. The limit curves are found to which
the eigenvalues concentrate and the counting eigenvalue functions along
these curves are obtained.
\vskip 2mm

\newpage


\begin{center}

{\LARGE\bf
О модельной задаче для уравнения Орра-Зоммерфельда \\[0.3cm]
c квадратичным профилем
}

\vspace{10mm}

{\large
C. Н. Туманов$^{(a)}$ и А. А. Шкаликов$^{(b)}$
}

\vspace{10mm}

Механико-математический факультет,\\
Московский Государственный Университет,\\
Москва, Россия\\
{\em

${}^{(a)}$e-mail: tumanov@art-com.ru\\
${}^{(b)}$e-mail: ashkalikov@yahoo.com\\
}

\vspace{15mm}

{\bf Аннотация}

\end{center}

\noindent
Рассматривается модельная задача для уравнения Орра--Зоммерфельда с 
произвольным квадратичным профилем. Полностью
изучено поведение спектра для больших чисел Рейнольдса. Найдены предельные
кривые, около которых концентрируются собственные значения, и получены
функции распределения собственных значений вдоль этих кривых.

\newpage

\section{Введение}

В работе изучается семейство дифференциальных операторов

\bgeq
\label{mainzad}
L(\varepsilon)y=i\varepsilon y''+q(x)y,\quad\varepsilon>0
\edeq
на вещественном отрезке $[a,b]$ с краевыми условиями:
\bgeq
\label{mainkraj}
y(a)=y(b)=0
\edeq
при $\varepsilon\to 0$.

Рассматриваемая задача является моделью для известного в
гидродинамике уравнения Орра--Зоммерфельда с профилем Пуазейля. Некоторые
аргументы, поясняющие, что спектральная 
задача~(\ref{mainzad}),~(\ref{mainkraj}) может служить
моделью для уравнения Орра--Зоммерфельда, приведены в~\cite{shmidt}. 
Однако, только в недавней работе автора и 
Шкаликова~\cite{tumscMatzams} дано строгое доказательство того, что 
спектральные портреты при $\varepsilon\to 0$ модельной и реальной 
задач совпадают. 

Число $\varepsilon>0$ играет роль малого параметра. Отметим, что 
$\varepsilon=1/R$, где~$R$~--- число Рейнольдса,
которое обратно пропорционально вязкости жидкости. Таким образом, малым
значениям параметра~$\varepsilon$ соответствует уравнение движения жидкости
близкой к идеальной.

Мы покажем, что при малых значениях $\varepsilon$ собственные значения
операторов~$L(\varepsilon)$ концентрируются около некоторых кривых в 
комплексной плоскости, вид которых укажем явно. Далее мы найдем асимптотические формулы
распределения собственных значений, равномерные по~$\varepsilon$, и укажем 
функции распределения собственных значений вдоль 
критических линий.

Подобная задача, когда вместо квадратичной функции в уравнении участвует
функция $q(x)=x$, изучена в работе~\cite{kuett}, где предельное множество, 
вдоль
которого концентрируются собственные значения, названо {\it спектральным
галстуком}. Граф критических линий в изучаемой задаче для 
$q(x)=\alpha x^2+\beta x+\gamma$ по форме
также напоминает галстук, но только несимметричный. Будем называть его
{\it предельным спектральным графом}.

Ранее в работе~\cite{petrovskij} задача~(\ref{mainzad}), (\ref{mainkraj})
была решена
для четного профиля Пуазейля ($q(x)=x^2$
с краевыми условиями Дирихле на отрезке $[-1,1]$).

Случай монотонного на отрезке профиля течения ранее
был рассмотрен в работе~\cite{kuett} А. А. Шкаликова. В ней доказывается,
что в этом особом случае предельное множество состоит из трех кривых. В связи
с этим, мы будем предполагать, что вершина соответствующей параболы $q(x)$
находится внутри рассматриваемого отрезка. Далее отметим, что случай 
принадлежности вершины центру отрезка заменой переменных приводится к случаю
$q(x)=x^2$ на $[-1,1]$, в связи с чем, в этой работе будем рассматривать 
только несимметричные профили. 

Дадим ряд важных определений и утверждений.

\begin{df}
\label{introdfone}
Точками поворота порядка $n$ функции $p$ называются ее нули. 
Точка поворота $z_0$ функции называется простой, если она является ее 
простым нулем. Кратностью точки поворота называется кратность ее как нуля
функции $p$.
\end{df}

Для любой точки поворота $z_0$ можно ввести многозначную функцию:
$$
S(z,z_0)=\int\limits_{z_0}^z\sqrt{p(\zeta)}\,d\zeta.
$$
\begin{df}
\label{introdf2}
Линией Стокса функции $p$, выходящей из точки поворота $z_0$,
называется кривая, для которой $$\re S(z,z_0)=0.$$ Кривая, определяемая
уравнением $\im S(z,z_0)=0$ называется сопряженной линией Стокса.
\end{df}

\begin{df}
\label{introdf3}
Комплексом Стокса функции $p$ называется всякое максимальное связное
множество, состоящее из линий Стокса. Комплекс Стокса называется простым, если
включает лишь одну простую точку поворота. В противном случае комплекс называется
сложным.
\end{df}

\begin{df}
Граф Стокса, состоящий из простых комплексов, называется простым. В противном
случае, граф называется сложным.
\end{df}

\begin{df}
Пара точек $a$ и $b$ имеет критическое расположение относительно 
простого комплекса Стокса, если одна из них лежит на линии Стокса,
а вторая находится либо на соседней линии Стокса, либо в области, ограниченной
соседними линиями Стокса, не содержащей линии Стокса, на которой лежит 
первая точка.
\end{df}

\begin{df}
\label{krittochkaigraphstoksa}
Пара точек $a$ и $b$ имеет критическое расположение относительно 
простого графа Стокса, если она имеет критическое расположение относительно
некоторого комплекса Стокса. В случае критического расположения та из точек
$a$ и $b$, которая лежит на линии Стокса называется критической, а число
$\la\in\Pi$, при котором точки расположены критически, называется сингулярным.
\end{df}

Отметим, что из определения простого комплекса Стокса следует

\begin{lm}
\label{intro_dvetreti}
Из всякой простой точки поворота выходят ровно три линии Стокса под углами
$\pm 2\pi/3$ по отношению друг к другу.
\end{lm}

Доказательство этого факта несложно и приводится в~\cite{fedor}.

\begin{df}
\label{introdf4}
Графом Стокса функции $p$ называется совокупность всех ее комплексов
Стокса.
\end{df}

Пусть функция $p(z,\la)$, зависит от параметра $\la$, рассмотрим семейство
ее графов Стокса $\Phi(\la)$. 

\begin{df}
Графы $\Phi(\la_1)$, $\Phi(\la_2)$ эквивалентны, если существует гомеоморфизм
$$
\phi:\Phi(\la_1)\to\Phi(\la_2),
$$ 
при котором точки поворота переходят в точки поворота.
\end{df}

\begin{df}
\label{introdopusttoch}
Точка $\la_0$ называется допустимой, если существует окрестность $O(\la_0)$ 
такая, что для всех точек $\la$ этой окрестности графы $\Phi(\la_0)$ и
$\Phi(\la)$ эквивалентны. В противном случае точка $\la_0$ называется
недопустимой.
\end{df}

Пусть 
$$
p(z,\la)=a_0(\la)(z-b_1(\la))^{n_1}\cdots (z-b_k(\la))^{n_k},\quad
n_j\in\N,
$$
где $b_j$ --- алгебраические функции, $a_0\not\equiv 0$ --- полином.

Положим
$$
\sa(a,b,\la)=\int\limits_a^b\sqrt{p(\zeta,\la)}\,d\zeta.
$$
Справедлива следующая лемма (см.~\cite{fedor} \S3 п.5):
\begin{lm}
\label{lm2introdukc}
Пусть $I_1$ --- нули $a_0$; $I_2$ --- множество всех $\la$: $b_j(\la)=b_l(\la)$,
$j\ne l$; $I_3$ --- множество всех $\la$, что при некоторых $j$, $l$ $j\ne l$
$\re \sa(b_j(\la),b_l(\la),\la)=0$. Тогда множество $I$ всех недопустимых точек
есть $I=I_1\cup I_2\cup I_3$.
\end{lm}

\begin{nt}
\label{lm3.2}
Для функции $p(z,\lambda)=i(z^2-\lambda)$ имеем
$$
I_1=\emptyset, \qquad
I_2=\{0\}, \qquad
I_3=\biggl\{\lambda\biggm|\arg\lambda=-\frac\pi4\biggr\},
$$
т.\,е.~ множество недопустимых точек в $\Pi$ совпадает с лучом
$\arg\lambda=-\pi/4$.
\end{nt}

\begin{df}
Максимальные области однолистности функции $S$, ограниченные линиями Стокса, 
содержащие точку поворота на своей границе, а внутри --- некоторую линию
Стокса, будем называть каноническими областями. (Смотри~\cite{fedor}).
\end{df}

\begin{df}
\label{kanon3-ka}
Тройку $(D_j,l_j,a_j)$, где $a_j$ --- точка поворота, $l_j$ --- линия Стокса,
выходящая из этой точки поворота, $D_j$ --- каноническая область, содержащая
$l_j$, будем называть канонической.
\end{df}

\begin{df}
\label{kanonsolution}
Пусть $(D_j,l_j,a_j)$ --- каноническая тройка. Ветвь функции $S(z,a_j)$, для
которой $\im S(l_j,a_j)=[0;+\infty)$, будем называть канонической ветвью, 
соответствующей данной тройке.
\end{df}

В конце работы мы приведем теорему о распределении собственных значений
изучаемой задачи. Она будет основана на следующем важном определении:

\begin{df}
\label{df5.4}
Функция $N(\la),$ определенная на 
$$
T=\tttttt\setminus\Bigl\{\la_j\Bigr\},
$$
где $\{\la_j\}$, ---
точки пересечения критических линий и конечные точки графа на вещественной 
оси, называется предельной 
спектральной функцией 
распределения, если для любой пары точек $\eta_1$, $\eta_2$, принадлежащей 
связной компоненте $T$, найдутся числа $C>0$ и $\varepsilon_0>0$ такие, 
что в окрестности 
$\Upsilon$ радиуса
$C\varepsilon$ интервала $(\eta_1,\eta_2)\subset T$ при 
$\varepsilon<\varepsilon_0$ 
число точек спектра будет равно
$$
{\cal N}
(\Upsilon)= |N(\eta_2)-N(\eta_1)|+O(1),\quad \varepsilon\to 0
$$
\end{df}

Здесь и далее будем обозначать
$$
p(x)=i(q(x)-\la).
$$      
\section{Постановка задачи и общие замечания о спектре}

Итак, рассмотрим краевую задачу~(\ref{mainzad}),~(\ref{mainkraj}).
Пусть
$$
q(x)=\alpha x^2+\beta x+\mu,\;\mbox{\rm где }\alpha\ne 0,\ \beta,\ \mu\in\R,
$$
наложим условие принадлежности вершины параболы внутренности отрезка:
$$
a<-\frac{\beta}{2\alpha}<b,
$$
наложим условие отсутствия четности параболы относительно це\-нтра отрезка:
$$
-\frac{\beta}{2\alpha}\ne\frac{a+b}{2}.
$$

Для произвольного $\theta\in\R$ cделаем замену переменной и параметров:
$$
z_1=\frac{1}{\theta}(z+\frac{\beta}{2\alpha}),
$$
$$
\la_1=\la+\frac{\beta^2}{4\alpha}-\mu,
$$
$$
u_1=|\theta|u,\quad u=\frac{1}{\sqrt{\varepsilon}}.
$$

В результате замены изучаемая спектральная задача примет вид:
$$
y''=iu_1^2(\alpha\theta^2z_1-\la_1)y
$$
$$
y(\frac{1}{\theta}(a+\frac{\beta}{2\alpha}))=
y(\frac{1}{\theta}(b+\frac{\beta}{2\alpha}))=0.
$$

Положим $\theta=\pm |\alpha|^{-1/2}$. За счет выбора знака можно добиться 
близости вершины параболы либо к левому, либо к правому концу отрезка.
Обратим внимание, что изменение знака $\alpha$ приводит к отражению спектра
относительно мнимой оси. Для доказательства этого факта достаточно рассмотреть
комплексное сопряжение уравнения и краевых условий.

В результате сформулируем следующее
\begin{nt}
\label{nt2.1m}
Задача~(\ref{mainzad}),~(\ref{mainkraj}) сдвигом переменной и 
сп\-ектрального параметра с возможным его последующих отражением относительно 
мнимой оси (если $\alpha<0$), растяжением $u$ приводится к виду:
$$
y''=iu^2(x^2-\la)y,
$$
$$
y(a_1)=y(b_1)=0,
$$
где $a_1<0<b_1$, $u>0$, а замена может быть выбрана так, чтобы
$|a_1|<|b_1|$.
\end{nt}

Основываясь на этом замечании, не ограничивая общности рассуждений, 
будем рассматривать следующую спектральную задачу:

\bgeq
\label{equiv2}
y''=iu^2(x^2-\la)y,\quad u>0,
\edeq
\bgeq
\label{cond2}
y(a)=y(b)=0,\quad a<0<b,\; |a|<|b|.
\edeq

В работе~\cite{petrovskij} была отмечена следующая, справедливая для 
вещественной непрерывной на отрезке функции $q$,

\begin{lm}
Спектр задачи~(\ref{mainzad}),~(\ref{mainkraj}) состоит из последовательности 
собственных значений $\la_n$, лежащих в полуполосе
$$
\Pi=\Bigl\{\la\in\C\Bigl|\min_{[a,b]}q(x)<\re\la<\max_{[a,b]}q(x),\ \im\la<0
\Bigr.\Bigr\},
$$
для которых существует предел вещественных частей:
$$
\lim_{n\to\infty}\re\la_n=\frac{1}{b-a}\int\limits_a^bq(x)\,dx.
$$
\end{lm}

\begin{nt}
Применительно к задаче~(\ref{equiv2}), (\ref{cond2})
$$
\Pi=\Bigl\{\la\in\C\Bigl|0<\re\la<b^2,\ \im\la<0
\Bigr.\Bigr\},
$$
$$
\lim_{n\to\infty}\re\la_n=\frac{a^2+ab+b^2}{3}.
$$
\end{nt}

\section{Граф Стокса}
\label{secgraphst}

С учетом замечания~\ref{nt2.1m} положим
$$
q(x)=x^2,\quad p(x)=i(q(x)-\la).
$$

Введем ветвь $\sqplus$, которую всюду далее будем называть {\it основной}:
$\arg\sqplus\in(-\pi/2,0)$ при $\im\lambda<0$. 

Приведенные ниже утверждения доказаны в работах~\cite{petrovskij} и~\cite{ts}.

\begin{lm}
\label{lm3.3}
Граф Стокса $\Gamma$ обладает свойством центральной симметрии относительно
точки~$0$.
\end{lm}

\begin{lm}
\label{lm3.4}
Граф $\Gamma(\lambda)$ связный тогда и только тогда, когда
$\psi=\arg\lambda=-\pi/4$.
\end{lm}

\begin{figure}

\caption{Внешние и внутренние линии графа Стокса.}
\label{vnvn}
\begin{center}

\end{center}

\end{figure}


\unitlength=35mm
\begin{figure}[ht]

\caption{Линии Стокса в случае $\arg\lambda\in(-\pi/4,0)$ 
(для $\lambda=1/2-i/4$).}
%
\label{part2ris2}
\end{figure}

\begin{figure}[ht]
\caption{Линии Стокса в случае $\arg\lambda=-\pi/4$ 
(для $\lambda=1/4-i/4$).}
%

\label{ris3}
\end{figure}

\begin{figure}[ht]
\caption{Линии Стокса в случае $\arg\lambda\in(-\pi/2,-\pi/4)$ 
(для $\lambda=1/4-i/2$).}
%
\label{ris4}
\end{figure}
\unitlength=1mm


\begin{df}
Пусть граф Стокса $i(z^2-\la)$ является простым. Тогда он состоит из двух 
простых комплексов Стокса. Линии Стокса каждого комплекса (по две на комплекс),
асимптотически
приближающиеся к линиям второго комплекса, называются внутренними. Остальные
линии
(по одной на комплекс) называются внешними (см. рис.~\ref{vnvn})
\end{df}

\begin{nt}
В дальнейшем важно понимать, какие области являются каноническими для
рассматриваемой функции~$p$. Рассмотрим два случая.

1) $\arg\lambda\ne-\pi/4$. Тогда граф~$\Gamma(\lambda)$ состоит из двух
комплексов Стокса (см.~рис.~\ref{part2ris2}, \ref{ris4}), которые разбивают плоскость на области
двух типов: типа полосы (область, ограниченная внутренними линиями Стокса) 
и типа полуплоскости (все остальные
области). В этом случае канонические области могут быть двух типов:

a) области, состоящие из двух областей типа полуплоскости и содержащие одну
внешнюю линию Стокса;

b) области, содержащие область типа полосы и две примыкающие к ней области
типа полуплоскости, отвечающие различным комплексам Стокса.

2) $\arg\lambda=-\pi/4$. Тогда граф Стокса состоит из одного комплекса,
разбивающего плоскость на четыре области типа полуплоскости 
(см. рис.~\ref{ris3}). В этом случае
канонические области состоят из любых двух соседних областей типа
полуплоскости.
\end{nt}

\begin{lm}
\label{simgrphz2}
Пусть $\la\in\Pi$, тогда граф Стокса функции $i(z^2-\la)$ обладает свойством 
центральной симметрии относите\-льно начала координат.
\end{lm}

\begin{lm}
\label{cerhbgfiv}
Пусть $\la\in\Pi$. Тогда изменение модуля $\la$ при фиксированном аргументе 
приводит к гомотетичному изменению графа Стокса.
\end{lm}

\begin{lm}
\label{grslozhtitd}
Пусть $-\pi/2<\arg\la<0$, тогда граф Стокса является сложным
тогда и только тогда, когда $\arg\la=-\pi/4$.
\end{lm}

\begin{lm}
\label{vnlinpervesos}
Пусть $-\pi/4<\arg\la<0$, тогда внешние линии Стокса функции $i(z^2-\la)$ 
пересекают вещественную ось.
\end{lm}

\begin{lm}
\label{vnlinnepervesos}
Пусть $-\pi/2<\arg\la<-\pi/4$, тогда внешние линии Стокса функции $i(z^2-\la)$ 
не пересекают вещественную ось.
\end{lm}

\section{Предельный спектральный граф}

В данном разделе мы докажем концентрацию спектра задачи 
(\ref{equiv2}), (\ref{cond2}) около предельного спектрального графа и явно
опишем составляющие его кривые.

вдоль предельных спектральных
кривых.
\begin{pr}
\label{print1}
Пусть $\xi>0$, тогда в области 
$$
K=\bigl\{
\la\in\C
\bigl|
\bigr.
\re\la>0,\
\im\la<0
\bigr\}
$$
при каждом $\psi=\arg\la$ найдется единственная точка кривой
$$
\tilde\gamma_\xi=\bigl\{
\la\in K
\bigl|
\bigr.
\re e^{\frac{\pi}{4}i}\int\limits_{\sqplus}^{\xi}\sqrt{\zeta^2-\la}\,d\zeta=0
\bigr\}.
$$

Более того, все точки кривой $\tilde\gamma_\xi$ лежат в полосе $0<\re\la<\xi^2$.
\end{pr} 

Введем кривую $\tilde\gamma_0$: $\arg\la=-\pi/4$. Она разбивает $\Pi$ на две 
подобласти: ограниченную 
и неограниченную. 
Ограниченную обозначим через $\Pi_r$, неограниченную --- через $\Pi_l$.
Рассмотрим две так называемых критических кривых (см.
рис.~\ref{pict1}):

$$
\tilde\gamma_a=\bigl\{
\la\in\Pi
\bigl|
\bigr.
\re e^{\frac{\pi}{4}i}\int\limits_{-\sqplus}^a\sqrt{\zeta^2-\la}\,d\zeta=0
\bigr\},
$$
$$
\tilde\gamma_b=\bigl\{
\la\in\Pi
\bigl|
\bigr.
\re e^{\frac{\pi}{4}i}\int\limits_{\sqplus}^b\sqrt{\zeta^2-\la}\,d\zeta=0
\bigr\}.
$$

\begin{figure}

\caption{Разбиение $\Pi$ критическими линиями.}
\label{pict1}
\begin{center}
\begin{picture}(60,60)

\put(10,0){\vector(0,1){55}}
\put(5,35){\vector(1,0){50}}
\put(40,0){\line(0,1){35}}
\put(10,35){\line(1,-1){30}}

\qbezier(40,35)(30,20)(10,15)
\qbezier(25,35)(20,30)(10,25)

\put(40,36){$b^2$}
\put(25,36){$a^2$}
\put(7,36){$0$}

\put(20,10){$\Pi_{1l}$}
\put(15,20){$\Pi_{2l}$}
\put(10.5,29){$\Pi_{3l}$}

\put(33,17){$\Pi_{1r}$}
\put(23,25){$\Pi_{2r}$}
\put(14.5,31){$\Pi_{3r}$}

\put(36,27){$\tilde\gamma_b$}
\put(24,32){$\tilde\gamma_a$}
\put(36,5){$\tilde\gamma_0$}

\end{picture}
\end{center}

\end{figure}

Согласно предложению~\ref{print1} и условию $a^2<b^2$, кривые расположены
именно так, как это показано на рисунке~\ref{pict1}. 
Вместе с $\tilde\gamma_0$ указанные
кривые разбивают полуполосу на шесть областей (см. рис.~\ref{pict1}). 

Через $\Pi_1$ обозначим объединение $\Pi_{1l}$ с $\Pi_{1r}$ и с частью 
$\tilde\gamma_0$
так, чтобы получилась одна из трех областей, на которые разбивается
$\Pi$ кривыми $\tilde\gamma_a$ и $\tilde\gamma_b$. Аналогично введем 
$\Pi_2$ и $\Pi_3$, содержащие в себе, соответственно, 
$\Pi_{2l(r)}$ и $\Pi_{3l(r)}$.

\begin{lm}
При всех $\la$ из $\Pi_{1}$, $\Pi_{2r}$, $\Pi_{3l}$, $\Pi_{3r}$ 
точки $a$, $b$ лежат в одной канонической области относительно гра\-фа Стокса.
\end{lm}
\prf 
Отметим, что всюду вне $\tilde\gamma_0$ графы Стокса являются простыми.
Для всех областей $\Pi_{jr}$ утверждение очевидно в силу того, что
в этих областях внешние линии пересекают граф Стокса 
(лемма~\ref{vnlinnepervesos}), стало быть, условие сингулярности задается
кривыми $\tilde\gamma_a$ и $\tilde\gamma_b$, находящимися на границе этих областей.
Далее, для всех $\la\in\Pi_{3l}$ ни одна из точек $a$ и $b$ не лежит в области
типа полосы графа Стокса, следовательно, сингулярными точками могут быть
лишь точки кривых $\tilde\gamma_a$ и $\tilde\gamma_b$. Следовательно, и в этой области нет
сингулярных точек. Аналогично доказывается отсутствие сингулярных точек в
$\Pi_{1l}$. Регулярность точек $\tilde\gamma_0\cap\Pi_1$ очевидна при учете 
лемм~\ref{cerhbgfiv},~\ref{grslozhtitd}.\edlm

\begin{lm}
В области $\Pi_{2l}$ существует кривая $\tilde\gamma_a^-$, задаваемая соотношением
$$
\re e^{\frac{\pi}{4}i}\int\limits_{\sqplus}^a\sqrt{\zeta^2-\la}\,d\zeta=0,
$$
вне которой для всех точек $\la\in\Pi_2\setminus\tilde\gamma_a^-$ 
точки $a$, $b$ лежат в одной канонической области относительно графа Стокса.
\end{lm}
\prf Действительно, при $\la\in\Pi_{2l}$ внешние линии Стокса не пересекают
вещественную ось, а точка $b$ находится в области типа полуплоскости 
относительно графа Стокса. Стало быть, принадлежность точки $a$ комплексу 
Стокса, содержащему точку поворота $\sqplus$, задает критическое расположение
пары точек $a$ и $b$ относительно графа Стокса с критической точкой $a$.
Всюду вне этой кривой утверждение леммы выполнено.\edlm

Отметим, что $\tilde\gamma_a^-$ одним из концов содержит точку $\la_1$ пересечения
$\tilde\gamma_a$ и $\tilde\gamma_0$. Можно показать, что $\tilde\gamma_a^-$ содержит и некоторую 
точку $\la_2\in\tilde\gamma_b$ (см. рис.~\ref{pict2}).


\begin{figure}

\caption{Критические линии.}
\label{pict2}
\begin{center}
\begin{picture}(60,60)

\put(10,0){\vector(0,1){55}}
\put(5,35){\vector(1,0){50}}
\put(40,0){\line(0,1){35}}
\put(10,35){\line(1,-1){30}}

\qbezier(40,35)(30,20)(10,15)
\qbezier(25,35)(20,30)(10,25)
\qbezier(16.5,28.5)(16,24)(20.5,18.5)

\put(40,36){$b^2$}
\put(25,36){$a^2$}
\put(7,36){$0$}

\put(36,27){$\tilde\gamma_b$}
\put(24,32){$\tilde\gamma_a$}
\put(36,5){$\tilde\gamma_0$}
\put(13,22){$\tilde\gamma_a^-$}
\put(15,30.5){$\la_1$}
\put(20,15.5){$\la_2$}

\end{picture}
\end{center}

\end{figure}

\begin{lm}
Точки $\la_1$ и $\la_2$ разбивают кривые $\tilde\gamma_a$ и $\tilde\gamma_b$ на две части.
При $\la$, лежащих в частях, примыкающих к мнимой оси, точки $a$, $b$ лежат в 
одной канонической области относительно графа Стокса.
\end{lm}
\prf Отметим, что кривая $\tilde\gamma_a^-$ разбивает $\Pi_{2l}$ на две области:
треугольник $T$ и четырехугольник $Q$. При $\la\in Q$ точка $a$ лежит в 
области типа полосы, а так как внешние линии не пересекают вещественной оси
при $\la\in\Pi_{2l}$, то выполнение условия принадлежности $b$ графу Стокса 
($\la\in\tilde\gamma_b$) не приведет к критическому расположению пары точек
$a$ и $b$. В случае с $\tilde\gamma_a$ критическому расположению пары мешает
принадлежность $b$ области типа полуплоскости комплекса, содержащего $\sqplus$.
\edlm

Из приведенных выше лемм вытекает следующее важное
\begin{cl}
Введем следующий набор кривых:
$$
\tapich=[0,\la_1],
$$
$$
\taa=
\bigl\{
\la\in\tilde\gamma_a
\bigl|
\bigr.
\arg\la\ge\arg\la_1
\bigr\},
$$
$$
\taam=\tilde\gamma_a^-,
$$
$$
\tab=
\bigl\{
\la\in\tilde\gamma_b
\bigl|
\bigr.
\arg\la\ge\arg\la_2
\bigr\}.
$$
Тогда при всех $\la\in\Pi$, не лежащих на указанных кривых 
точки $a$, $b$ лежат в одной канонической области относительно гра\-фа Стокса.
(см. рис.~\ref{pict3}).
\end{cl}

\begin{figure}

\caption{Критическое расположение пары.}
\label{pict3}
\begin{center}
\begin{picture}(60,60)

\put(10,0){\vector(0,1){55}}
\put(5,35){\vector(1,0){50}}
\put(40,0){\line(0,1){35}}
\put(10,35){\line(1,-1){6.5}}

\qbezier(40,35)(30,20)(20.5,18.5)
\qbezier(25,35)(20,30)(16.5,28.5)
\qbezier(16.5,28.5)(16,24)(20.5,18.5)

\put(40,36){$b^2$}
\put(25,36){$a^2$}
\put(7,36){$0$}

\put(35,25){$\tab$}
\put(23,30){$\taa$}
\put(11,28){$\tapich$}
\put(12,22){$\taam$}
\put(15,30.5){$\la_1$}
\put(20,15.5){$\la_2$}

\end{picture}
\end{center}

\end{figure}

Отметим

\begin{pr}
При $\la\in\Pi\setminus\Pi_{1l}$ точки $a$ и $b$ не могут находиться 
в области типа полосы одновременно.
\end{pr}

Рассмотрим функции
$$
\tilde w_1(z,\lambda,u)
=p^{-1/4}e^{+uS(z,\sqplus)},
$$
$$
\tilde w_2(z,\lambda,u)
=p^{-1/4}e^{-uS(z,\sqplus)}.
$$

Функции $\tilde w_1$, \ $\tilde w_2$  известны как ВКБ-приближения решений
уравнения~(\ref{equiv2}). 
Следующие две леммы доказаны в~\cite[\S\,3, п.~5]{fedor}.

\begin{lm}
\label{lm4.1}
Пусть $\Omega$~--- связный компакт в $\lambda$-плоскости, все точки которого
допустимы в силу определения~\ref{introdopusttoch}
(в нашем случае в силу замечания~\ref{lm3.2} компакт~$\Omega$ не содержит
точек~$\lambda$: $\arg\lambda=-\pi/4$). Пусть~$D(\lambda)$~--- канонические
области функции~$p(z,\lambda)$, непрерывно зависящие от $\lambda\in\Omega$.
Тогда существует фундаментальная система решений~$w_1$, \ $w_2$
уравнения~(\ref{equiv2}), для которой при
$z\in D=\cap_{\lambda\in K}D(\lambda)$ справедливы асимптотики
\bgeq
\label{4.3}
w_{1,2}(z,\lambda,u)=
\tilde w_{1,2}(z,\lambda,u)\biggl(1+O\biggl(\frac1u\biggr)\biggr).
\edeq
Указанные асимптотики можно дифференцировать по $z$, $u$, $\lambda$ любое
число раз.
\end{lm}

Если $\lambda_0$~--- недопустимая точка, то может существовать семейство
канонических областей~$D(\lambda)$, непрерывно зависящее от~$\lambda$ при
$\lambda\not\in I_3$, которое при $\lambda\to\lambda_0$ имеет пределом
неодносвязную область, следовательно, не каноническую. Но могут также
существовать области~$D(\lambda)$, которые остаются непрерывными при
~$\lambda$, принадлежащих всему множеству~$I_3$ или его части. В такой
ситуации лемма~\ref{lm4.1} допускает обобщение.

\begin{lm}
\label{lm4.2}
Пусть $\Omega$~--- связный компакт в $\lambda$-плоскости, все точки которого
либо допустимы, либо принадлежат типу~$I_3$ (для рассматриваемого случая
функции $q=x^2$ область~$\Omega$ не содержит~$0$). Пусть канонические
области~$D(\lambda)$ таковы, что при $\lambda\in\Omega$ они непрерывны. Тогда
справедливо утверждение леммы~\ref{lm4.1}.
\end{lm}

В работе~\cite{ts} доказана равномерность асимптотических соотношений~(\ref{4.3})
как при большом параметре $u$, так и при большом спектральном параметре
$\la$. Здесь мы приведем формулировку соответствующей леммы.

\begin{lm}
\label{lmravnomula}
Для любого $K$ --- компакта в $\C$ и $u_0>0$ найдется число $R_0>0$ такое, что 
уравнение~(\ref{equiv2}) будет иметь пару независимых решений:
$$
U(z)= cp^{-1/4}e^{u S(z,\sqplus)}(1+\frac{\epsilon_+(z,u,\la)}{u\la}),\quad
$$
$$
V(z)= cp^{-1/4}e^{-u S(z,\sqplus)}(1+\frac{\epsilon_-(z,u,\la)}{u\la}),
$$
где как и ранее
$$
S(z,\sqplus)=\sa(\sqplus,z,\la)=
e^{\frac{\pi}{4}i}\int\limits_{\sqplus}^z\sqrt{\zeta^2-\la}\,d\zeta,
\quad p(z)=i(z^2-1),
$$
а $\epsilon_{\pm}(z,u,\la)$ --- ограниченные функции при 
$z\in K$ и $|\la|>R_0$, $u>u_0$.
\end{lm}

\begin{lm}
\label{vrvrgvunrnhavk}
Пусть для всех $\la$ из некоторого замкнутого подмножества $G\subset\Pi$ 
точки $a$ и $b$ 
не расположены критически относительно простого графа Стокса. В этом случае
либо найдется $u_0>0$ такое, что при $u>u_0$ множество $G$ не будет содержать
собственных значений задачи~(\ref{equiv2}),~(\ref{cond2}), либо множество
\bgeq
\label{osneqnakrivvhorobl}
\tau=\bigl\{
\la\in G
\bigl|
\bigr.
\re\sa(\sqplus,a,\la)=
\re\sa(\sqplus,b,\la)
\bigr\}
\edeq
не пусто. В последнем случае для всякого $\delta>0$ найдется
$u_0>0$ такое, что при $u>u_0$ вне $\delta$-окрестности $\tau$
в $G$ не будет содержаться
собственных значений задачи~(\ref{equiv2}),~(\ref{cond2})
\end{lm}
\prf Обратим внимание, что те и только те $\la$, при которых граф Стокса
является простым, являются допустимыми, в связи с чем в рассматриваемом
множестве $G$ можно применять леммы данного раздела. 

После вычисления характеристического определителя с последующим его
приравниванием к нулю, получим уравнение на $\tau$. 

Обратим внимание, что всюду в $G$ мы можем пользоваться аси\-мптотиками 
решений $w_1$ и $w_2$, так как всюду в $G$ точки $a$ и $b$ не расположены
критически относительно графа Стокса.

Более подробное доказательство было проведено для четного профиля в 
работе~\cite{petrovskij}.\edlm

\begin{thrma}
\label{mnth3chapt1}
Рассмотрим следующие кривые:
$$
\tapich=[0,\la_1],
$$
$$
\taa=
\bigl\{
\la\in\tilde\gamma_a
\bigl|
\bigr.
\arg\la\ge\arg\la_1
\bigr\},
$$
$$
\taam=\tilde\gamma_a^-,
$$
$$
\tab=
\bigl\{
\la\in\tilde\gamma_b
\bigl|
\bigr.
\arg\la\ge\arg\la_2
\bigr\},
$$
кроме того, введем
$$
\tainf=
\bigl\{
\la\in\Pi_{1l}
\bigl|
\bigr.
\re e^{\frac{\pi}{4}i}\int\limits_a^b\sqrt{\zeta^2-\la}\,d\zeta=0
\bigr\}.
$$
Обозначим (см. рис~\ref{pict4})
$$
\tttttt=\tapich\cup\taa\cup\taam\cup\tab\cup\tainf.
$$

Тогда для всякой $\delta$-окрестности $\tttttt$ 
найдется $u_0>0$ такое, что
при $u>u_0$ все точки спектра будут содержаться в этой окрестности.
\end{thrma}

\begin{figure}

\caption{Предельные спектральные кривые.}
\label{pict4}
\begin{center}
\begin{picture}(60,60)

\put(10,0){\vector(0,1){55}}
\put(5,35){\vector(1,0){50}}
\put(40,0){\line(0,1){35}}
\put(10,35){\line(1,-1){6.5}}

\qbezier(40,35)(30,20)(20.5,18.5)
\qbezier(25,35)(20,30)(16.5,28.5)
\qbezier(16.5,28.5)(16,24)(20.5,18.5)
\qbezier(20.5,18.5)(22,10)(22,0)

\put(40,36){$b^2$}
\put(25,36){$a^2$}
\put(7,36){$0$}

\put(35,25){$\tab$}
\put(23,30){$\taa$}
\put(11,28){$\tapich$}
\put(12,22){$\taam$}
\put(22.5,10){$\tainf$}
\put(15,30.5){$\la_1$}
\put(20,20.5){$\la_2$}

\end{picture}
\end{center}

\end{figure}

\prf Покажем, что в областях $\Pi_{1r}$, $\Pi_{2l}$,
$\Pi_{2r}$, $\Pi_{3l}$, $\Pi_{3r}$ нет точек, множества
$\tau$~(\ref{osneqnakrivvhorobl}).

Сначала рассмотрим $\Pi_{1r}$, $\Pi_{2r}$, $\Pi_{3r}$. Указанные области
лежат в угле $-\pi/4<\arg z <0$, следовательно, согласно 
лемме~\ref{vnlinpervesos} для всех точек
$\la$ из этих областей внешние линии Стокса пересекают вещественную ось.
Отметим, что при этом ни одна из точек $a$ и $b$ не может находиться в области
типа полосы (ограниченной внутренними линиями Стокса), иначе бы существовала
пара точек $x_{1,2}^+>b$ (в случае принадлежности $b$ области типа полосы),
или $x_{1,2}^-<a$ (в случае принадлежности $a$ области типа полосы) --- точки 
пересечения комплекса Стокса с вещественной осью, из чего
следовало бы противоречие с единственностью точки кривой
$\tilde\gamma_a$ или с единственностью точки кривой $\tilde\gamma_b$ при фиксированном 
аргументе. Стало быть, между точками $a$ и $b$ проходит линия Стокса 
комплекса Стокса, выходящего из $\sqplus$, следовательно, знаки вещественных
частей различны:
$$
\sgn\re\sa(\sqplus,a,\la)=-
\sgn\re\sa(\sqplus,b,\la)
$$
при $\la\in\Pi_{1r}\cup\Pi_{2r}\cup\Pi_{3r}$.

Рассмотрим $\Pi_{2l}$ и $\Pi_{3l}$. Отметим, что в этих областях точка
$b$ не лежит в области типа полосы, следовательно, между парой точек 
$a$ и $b$ снова лежит линия Стокса комплекса, выходящего из $\sqplus$ и
знаки вещественных частей различаются.

Обратим внимание, что в области $\Pi_{1l}$ можно применить 
лемму \ref{vnlinnepervesos} и выбрать ветвь $\sqrt{\zeta^2-\la}$
с разрезами по внешним линиям Сто\-кса, которые не пересекут вещественной оси,
в связи с чем уравнение~(\ref{osneqnakrivvhorobl}) на $\tau$ можно заменить
интегралом, определяющим $\tainf$.

Дальнейшее очевидное применение леммы~\ref{vrvrgvunrnhavk} приводит к 
завершению доказательства.\edth

Обратим внимание, что все критические кривые можно разбить на кривые трех 
типов:
\begin{enumerate}
\item\label{firstcurve}
Характеризуется тем, что одна из точек $a$ или $b$ находится на линии Стокса
простого комплекса, а вторая --- в той части комплексной плоскости,
разбиваемой данным комплексом на три части, которая не содержит эту
линию Стокса. В нашем случае такими являются кривые: $\taa$, $\taam$,
$\tab$.
\item\label{secondcurve}
Для всех точек $\la$ кривой данного типа графы Стокса являются сложными,
а пара точек $a$ и $b$ не лежит в одной канонической области относительно 
графа. В нашем случае такой кривой является $\tapich$.
\item\label{thirdcurve}
Характеризуется тем, что для всех точек $\la$ такой кривой граф Стокса
является простым, пара точек $a$ и $b$ лежит в одной канонической области,
а сама кривая удовлетворяет уравнению~(\ref{osneqnakrivvhorobl}) на $\tau$.
Применительно к рассматриваемому случаю такой является кривая $\tainf$.
\end{enumerate}

\section{Поведение спектра при $\varepsilon\to 0$}

Нам потребуются утверждения о матрицах перехода, доказательства которых
можно найти в монографии~\cite{fedor}.

Пусть даны две пары $(U_j,V_j)$, \ $(U_k,V_k)$ линейно независимых решений
уравнения~(\ref{equiv2}). Здесь индексы~$j$ и~$k$ фиксированы. Произвольное
решение~$W$ этого уравнения можно представить линейной комбинацией этих пар
решений:
$$
W=\alpha_jU_j+\beta_jV_j=\alpha_kU_k+\beta_kV_k.
$$

\begin{df}
\label{df5.1}
Матрица $\Omega_{jk}=\Omega_{jk}(u,\lambda)$ такая, что
$$
\left(
\begin{array}{c}
\alpha_k
\\
\beta_k
\end{array}
\right)
=\Omega_{jk}
\left(
\begin{array}{c}
\alpha_j
\\
\beta_j
\end{array}
\right)
$$
для всех решений $W$ уравнения~(\ref{equiv2}), называется {\it матрицей
перехода от пары $(U_j,V_j)$ к паре\/} $(U_k,V_k)$.  В частности, при всех
$z\in\C$ справедливо равенство
$$
\left(
\begin{array}{c}
U_j
\\
V_j
\end{array}
\right)
=\Omega_{jk}^T
\left(
\begin{array}{c}
U_k
\\
V_k
\end{array}\right),
$$
где $\Omega_{jk}^T$ --- транспонированная матрица перехода.
\end{df}

Как и ранее, положим $p(z,\lambda)=i(z^2-\lambda)$. Рассмотрим каноническую
тройку $(D,l,z_0)$ и каноническую ветвь~$S$ (в смысле
определений~\ref{kanon3-ka},~\ref{kanonsolution}). Удалим из~$S(D)$ левые (правые) $\varepsilon$-окрестности
разрезов и $\varepsilon$-окрестности образов точек поворота. Прообраз обозначим через
~$D_\varepsilon^+$ ($D_\varepsilon^-$).

\begin{df}
\label{df5.2}
Бесконечный путь $\gamma^+(z)\subset D^+_\varepsilon$
($\gamma^-(z)\subset D^-_\varepsilon$)
называется {\it положительным\/} ({\it отрицательным\/}) {\it
каноническим путем для тройки\/} $(D,l,z_0)$, если один из его концов
совпадает с~$z$ и для канонической ветви~$S$ соотношение 
$\re S(\zeta,z_0)\to+\infty(-\infty)$ имеет место при $\zeta\in \gamma^+(z)$
($\zeta\in\gamma^-(z)$), $z\to\infty$.
\end{df}

\begin{lm}
\label{lm5.1}
Пусть $(D,l,z_0)$~--- каноническая тройка, $S$~--- каноническая ветвь.
Тогда при $u\gg1$ в области~$D$ существует пара независимых решений~$(U,V)$,
именуемая канонической парой, соответствующей данной канонической тройке,
асимптотически представляемых в~$D$ разложениями при $u\to+\infty$:
$$
U(z,u)
\sim cp^{-1/4}(z)e^{+uS(z,z_0)}\exp\biggl(\,\sum_{k=1}^\infty
(+u)^{-k}\int_{\gamma^+(z)}\alpha_k(t)\,dt\biggr),
$$
$$
V(z,u)
\sim cp^{-1/4}(z)e^{-uS(z,z_0)}\exp\biggl(\,\sum_{k=1}^\infty
(-u)^{-k}\int_{\gamma^-(z)}\alpha_k(t)\,dt\biggr),
$$
где
$$
\alpha_1=\frac18\frac{p''}{p^{3/2}}-\frac{5}{32}\frac{(p')^2}{p^{5/2}}, \qquad
\alpha_{k+1}=-\frac1{2\sqrt{q}}
\biggl(\alpha_k'+\sum_{j=0}^k\alpha_j\alpha_{k-j}\biggr)
$$
и для удобства записи используются обозначения
$$
\alpha_{-1}=\sqrt p, \qquad
\alpha_0=-\frac{p'}{4p}.
$$
Здесь $\gamma^\pm(z)$~--- произвольные положительный и отрицательный
канонические пути в~$D$, а~$c$~--- нормировочный коэффициент такой, что
$$
|c|=1, \qquad
\lim_{z\to z_0,z\in l}\arg\bigl(cp^{-1/4}(z)\bigr)=0.
$$
\end{lm}
\begin{nt}
\label{nt5.1} 
Существование бесконечного числа членов
в приведенных асимптотических разложениях можно доказать,
в частности, для полиномиальных функций $q$. Однако,
далее используется только первое приближение.
\end{nt}

\begin{df}
Пусть заданы две канонические тройки $(D_j,l_j,z_j)$ и
$(D_k,\allowbreak l_k,z_k)$.
Матрицу
перехода от канонической пары решений $(U_j,V_j)$ к канонической паре 
$(U_k,V_k)$ будем называть {\it матрицей перехода
от канонической тройки $(D_j,l_j,z_j)$ к $(D_k,l_k,z_k)$}.
\end{df}

Можно доказать существование четырех типов матриц перехода между
каноническими тройками, из которых путем композиций получаются все другие
возможные матрицы перехода. Сформулируем леммы, представляющие вид матриц в
трех специальных ситуациях.

\begin{lm}
\label{lm5.2}
Пусть $(D,l,z_1)$~--- первая тройка, $(D,l,z_2)$~--- вторая. Меняется лишь
направление конечной линии Стокса~$l$. Тогда
$$
\Omega=e^{i\varphi_0}
\left(\begin{array}{cc}
0 & e^{-iua}
\\
e^{iua} & 0
\end{array}\right), \qquad
a=|S(z_1,z_2)|, \quad
e^{i\varphi_0}=c_2/c_1.
$$
\end{lm}

\begin{lm}
\label{lm5.3}
Пусть $(D,l_1,z_1)$~--- первая тройка, $(D,l_2,z_2)$~--- вторая, причем
лучи~$S(l_1)$ и~$S(l_2)$ направлены в одну сторону. Пусть~$l_2$ находится
слева от~$l_1$. Тогда \ $a=S(z_1,z_2)$, $\re a>0$ и
$$
\Omega=e^{i\varphi_0}
\left(\begin{array}{cc}
e^{-ua} & 0
\\
0 & e^{ua}
\end{array}\right), \qquad
e^{i\varphi_0}=c_2/c_1.
$$
\end{lm}

\begin{lm}
\label{lm5.4}
Пусть $z_0$~--- простая точка поворота (нуль первого порядка функции~$p$),
$l_1$, \ $l_2$, \ $l_3$~--- линии Стокса с началом~$z_0$, \ $l_{j+1}$ лежит
слева от~$l_j$.
Обозначим матрицу
перехода от $(D_j,l_j,z_0)$ к $(D_{j+1},l_{j+1},z_0)$
через~$\Omega_{j,j+1}$.  Тогда 
$$ 
\Omega_{j,j+1}=e^{-\frac\pi6i}
\left(\begin{array}{cc}
0 & \alpha_{j,j+1}^{-1} \\ 1 & i\alpha_{j+1,j+2} 
\end{array}\right), \qquad
\alpha_{12}\alpha_{23}\alpha_{31}=1,
$$
и справедливо асимптотическое разложение
\bgeq
\label{5.1}
\alpha_{j,j+1}\asymp\exp\biggl(\,\sum_{k=1}^\infty
(-u)^{-k}\int_{\gamma_{j,j+1}}\alpha_k(t)\,dt\biggr).
\edeq
Здесь бесконечный контур $\gamma_{j,j+1}$ лежит в $D_j\cup D_{j+1}$,
начинается в~$D_{j+1}$ там, где $\re S\to+\infty$, и заканчивается в~$D_j$
там, где $\re S\to -\infty$. Ветвь~$\sqrt p$ выбрана так же, как для
фундаментальной системы~$(U_j,V_j)$.
\end{lm}

\begin{figure}
\label{picrpa}
\caption{Матрицы перехода и канонические области.}
\begin{center}
\begin{picture}(145,80)

\kan
\put(5,55){\line(1,2){5}} 
\nan
\put(5,75){\line(1,-2){5}}
\put(30,65){\line(1,2){5}}
\put(30,65){\line(1,-2){5}}
\qbezier(10,65)(30,60)(33,55)
\qbezier(30,65)(10,70)(7,75)

\put(40,55){\line(1,2){5}} 
\put(40,75){\line(1,-2){5}}
\put(65,65){\line(1,2){5}}
\put(65,65){\line(1,-2){5}}
\kan
\qbezier(45,65)(65,60)(68,55) 
\nan
\qbezier(65,65)(45,70)(42,75)

\put(75,55){\line(1,2){5}} 
\put(75,75){\line(1,-2){5}}
\put(100,65){\line(1,2){5}}
\put(100,65){\line(1,-2){5}}
\qbezier(80,65)(100,60)(103,55) 
\kan
\qbezier(100,65)(80,70)(77,75) 
\nan

\put(110,55){\line(1,2){5}} 
\put(110,75){\line(1,-2){5}}
\kan
\put(135,65){\line(1,2){5}} 
\nan
\put(135,65){\line(1,-2){5}}
\qbezier(115,65)(135,60)(138,55) 
\qbezier(135,65)(115,70)(112,75) 

\put(5,50){\line(1,-2){5}}
\kan
\put(5,30){\line(1,2){5}} 
\nan
\put(10,40){\line(1,0){20}}
\put(30,40){\line(1,2){5}}
\put(30,40){\line(1,-2){5}}

\put(40,50){\line(1,-2){5}}
\put(40,30){\line(1,2){5}} 
\kan
\put(45,40){\line(1,0){20}} 
\nan
\put(65,40){\line(1,2){5}}
\put(65,40){\line(1,-2){5}}

\put(75,50){\line(1,-2){5}}
\put(75,30){\line(1,2){5}} 
\kan
\put(80,40){\line(1,0){20}} 
\nan
\put(100,40){\line(1,2){5}}
\put(100,40){\line(1,-2){5}}

\put(110,50){\line(1,-2){5}}
\put(110,30){\line(1,2){5}} 
\put(115,40){\line(1,0){20}} 
\kan
\put(135,40){\line(1,2){5}} 
\nan
\put(135,40){\line(1,-2){5}}

\put(5,25){\line(1,-2){5}}
\kan
\put(5,5){\line(1,2){5}} 
\nan
\put(30,15){\line(1,2){5}}
\put(30,15){\line(1,-2){5}}
\qbezier(10,15)(30,20)(33,25)
\qbezier(30,15)(10,10)(7,5)

\put(40,25){\line(1,-2){5}}
\put(40,5){\line(1,2){5}} 
\put(65,15){\line(1,2){5}}
\put(65,15){\line(1,-2){5}}
\kan
\qbezier(45,15)(65,20)(68,25) 
\nan
\qbezier(65,15)(45,10)(42,5)

\put(75,25){\line(1,-2){5}}
\put(75,5){\line(1,2){5}} 
\put(100,15){\line(1,2){5}}
\put(100,15){\line(1,-2){5}}
\qbezier(80,15)(100,20)(103,25) 
\kan
\qbezier(100,15)(80,10)(77,5) 
\nan

\put(110,25){\line(1,-2){5}}
\put(110,5){\line(1,2){5}} 
\kan
\put(135,15){\line(1,2){5}} 
\nan
\put(135,15){\line(1,-2){5}}
\qbezier(115,15)(135,20)(138,25) 
\qbezier(135,15)(115,10)(112,5) 

\put(10,65){\circle*{1.5}}
\put(45,65){\circle*{1.5}}
\put(100,65){\circle*{1.5}}
\put(135,65){\circle*{1.5}}

\put(10,40){\circle*{1.5}}
\put(45,40){\circle*{1.5}}
\put(100,40){\circle*{1.5}}
\put(135,40){\circle*{1.5}}

\put(10,15){\circle*{1.5}}
\put(45,15){\circle*{1.5}}
\put(100,15){\circle*{1.5}}
\put(135,15){\circle*{1.5}}

\put(7,29){$l_1$}
\put(15,36){$l_2$}
\put(50,36){$l_2$}
\put(85,36){$l_4$}
\put(127,36){$l_4$}
\put(142,29){$l_5$}
\put(7,49){$l_3$}
\put(142,49){$l_6$}

\put(2,4){$l_1$}
\put(7,54){$l_1$}
\put(142,24){$l_6$}
\put(142,74){$l_6$}

\put(20,32){$D_1$}
\put(4,39){$D_1$}
\put(55,32){$D_2$}
\put(48,45){$D_2$}
\put(90,32){$D_4$}
\put(83,45){$D_4$}
\put(118,45){$D_6$}
\put(137,39){$D_6$}

\put(35,39){$\Omega_{12}$}
\put(35,36){$\longrightarrow$}
\put(70,39){$\Omega_{24}$}
\put(70,36){$\longrightarrow$}
\put(105,39){$\Omega_{46}$}
\put(105,36){$\longrightarrow$}

\put(14,75){$\arg\la>-\frac{\pi}{4}$}
\put(14,50){$\arg\la=-\frac{\pi}{4}$}
\put(14,25){$\arg\la<-\frac{\pi}{4}$}

\end{picture}
\end{center}

\end{figure}

Исследования спектра на различных кривых предельного спектрального графа
технически схожи, поэтому, более подробно рассмотрим наиболее сложный случай
--- кривой типа~\ref{secondcurve} (случай $\gamma_0$).

При движении вдоль кривой,
пересекающей $\tapich$, вдоль которой аргумент строго убывает, граф
претерпевает изменения, показанные на рис.~\ref{picrpa}. Верхний ряд
соответствует тем точкам кривой, для которых $\arg\la> -\pi/4$; средний ряд
--- $\arg\la =-\pi/4$; нижний --- $\arg\la< -\pi/4$. 

Обозначим линии Стокса и канонические тройки так, как это сделано на
рисунке~\ref{picrpa}. Индексы канонических решений положим равными
индексам канонических областей.

\begin{lm}
\label{omega21javno}
Матрица $\Omega_{16}$ перехода от канонической тройки $(D_1,l_1,-\sqplus)$ к тройке
$(D_6,l_6,\sqplus)$ выражается следующей фо\-рмулой:
$$
\Omega_{16}^T=\left(
\begin{array}{cc}
\omega_{11} & \omega_{12}\\
\omega_{21} & \omega_{22}
\end{array}
\right)=
C\left(
\begin{array}{cc}
-i\alpha_{45}\alpha_{64}e^{-iud} & \alpha_{64}e^{-iud}\\
\alpha_{23}\alpha_{45}\alpha_{64}e^{-iud}+\frac{1}{\alpha_{12}}e^{iud} &
i\alpha_{23}\alpha_{64}e^{-iud}
\end{array}
\right),
$$
$$
\quad d=|\la|\frac{\pi}{2}.
$$
\end{lm}
\prf Из обозначений для матриц, введенных на рисунке~\ref{picrpa} следует,
что
$$
\Omega_{16}^T=\Omega_{12}^T\Omega_{24}^T\Omega_{46}^T=
\Omega_{12}^T\Omega_{24}^T(\Omega_{64}^T)^{-1},
$$
последние матрицы нам даются своими явными видами леммами \ref{lm5.2}
и \ref{lm5.4}. 
$$
d=\left|e^{\frac{\pi}{4}i}\la\int\limits_{-1}^1\sqrt{\zeta^2-1}\,d\zeta
\right|=
|\la|\frac{\pi}{2}.
$$
\edlm

Рассмотрим характеристический определитель в некоторой ок\-рестности $\Upsilon$
интервала в $\tapich$ в $\Pi$, не содержащей некоторой окрестности нуля.
$$
\Delta(u,\la)=
\left|
\begin{array}{cc}
U_1(a) & U_1(b)\\
V_1(a) & V_1(b)
\end{array}
\right|=\left|
\begin{array}{cc}
U_1(a) & \omega_{11}U_6(b)+\omega_{12}V_6(b)\\
V_1(a) & \omega_{21}U_6(b)+\omega_{22}V_6(b)
\end{array}
\right|=
$$
$$
=\omega_{21}U_1(a)U_6(b)+\omega_{22}U_1(a)V_6(b)-
\omega_{11}V_1(a)U_6(b)-\omega_{12}V_1(a)V_6(b).
$$

Рассмотрим канонические ветви $S_6(-\sqplus,b)$ и $S_1(\sqplus,a)$. 

\begin{lm}
\label{talI0la0}
Для любого отрезка $I$ множества $\tapich$, не содержащего точек $0$ и $\la_1$,
найдется окрестность $\Upsilon$ этого отрезка, для всех точек $\la$ которой
$$
\re S_1(-\sqplus,a)>0,\quad\re S_6(\sqplus,b)>0.
$$
\end{lm}
\prf Немедленно следует из расположения точек $a$ и $b$ относительно графа 
Стокса при $\la\in\Upsilon$ и аналитичности функций.
\edlm
\begin{lm}
\label{thtalUpsk}
Для любой точки $\mu\in\tapich$ такой, что $\mu\ne 0$ и $\mu\ne\la_1$, для любой 
окрестности $O$
этой точки, не содержащей некоторых окрестностей нуля и $\la_1$,
найдутся $u_0>0$ и число $C>0$ такие, что при всех $u>u_0$ окрестность $O$
будет содержать точки спектра.
Более того, если рассмотреть точки
$\tilde\la_k=e^{-i\pi/4}(2k+1)/u$,
то в каждой окрестности ${\cal U}_k$ этих точек
радиуса $C/u^2$, лежащей в $O$,
найдется и при том единственная точка спектра.
\end{lm}
\prf Рассмотрим окрестность $\Upsilon$ произвольного отрезка в $\tapich$,
содержащего $\mu$, которую нам дает лемма~\ref{talI0la0}. Не ограничивая 
общности, будем считать $O\subset\Upsilon$.

Пользуясь леммой~\ref{talI0la0}, получим, что при $u\to +\infty$ равномерно в
$O$ справедливо асимптотическое равенство:
$$
\Delta(u,\la)=\omega_{21}p^{-1/4}(a)p^{-1/4}(b)e^{u(S_6(b)+S_1(a))}
(
1+o(e^{-u\delta})
).
$$

Приравняем характеристический определитель к нулю, получим эквивалентное
уравнение, справедливое при $u\gg 1$ равномерно в $O$:
$$
\omega_{21}=0,
$$
пользуясь явным видом данного элемента (см. лемму~\ref{omega21javno}), получим
$$
\frac{\alpha_{23}}{\alpha_{56}}e^{-iud}+\frac{1}{\alpha_{12}}e^{iud}=0
\;\Leftrightarrow\;
e^{2iud}=-\frac{\alpha_{31}}{\alpha_{56}},
$$
или
\bgeq
\label{fndlatal}
e^{\pi iue^{i\pi/4}\la}=-1+O(\frac{1}{u}),
\edeq
где функция, являющаяся равномерным $O(1/u)$, аналитическая в окрестности $O$.
Далее воспользуемся теоремой Руше.
\edlm

Данная лемма очевидным образом обобщается для всякого отрезка $I\subset\tapich$,
не содержащего точек $0$ и $\la_1$.
\begin{cl}
\label{cltalUpsk}
Для любого отрезка $I\subset\tapich$ $0\not\in I$ и $\la_1\not\in I$,
для любой окрестности $\Upsilon$ этого отрезка, не содержащей некоторых окрестностей
нуля и $\la_1$,
найдутся $u_0>0$ и число $C>0$ такие, что при всех $u>u_0$ окрестность
$\Upsilon$ будет содержать точки спектра.
Более того, если рассмотреть точки
$\tilde\la_k=e^{-i\pi/4}(2k+1)/u$, то в каждой окрестности
${\cal U}_k$ этих точек
радиуса $C/u^2$, лежащей в $\Upsilon$
найдется и при том единственная точка спектра.
\end{cl}

Следующее утверждение дает нам распределение спектра в окрестности кривых 
$\taa$, $\taam$, $\tab$.

\begin{lm}
\label{lm2ospectre}
Пусть кривая $\gamma$ характеризуется тем, что из пары точек $a$, $b$ найдется
одна (обозначим ее через $\xi$), являющаяся при всех $\la\in\gamma$ 
критической в смысле определения~\ref{krittochkaigraphstoksa}. Обозначим
$z_0(\la)$ --- точка поворота, из которой исходит линия Стокса содержащая
$\xi$. Тогда для любого интревала $I\subset\gamma$, не содержащего концевых
точек $\gamma$, для любой окрестности $\Upsilon$ этого интервала, не содержащей
некоторых окрестностей концевых точек $\gamma$, найдутся $u_0>0$ 
и число $C>0$ такие, что при всех $u>u_0$ окрестность
$\Upsilon$ будет содержать точки спектра.
Более того, если рассмотреть точки $\tilde\la_k$, удовлетворяющие соотношению
$\sa(\tilde\la_k,z_0(\tilde\la_k),\xi)
=(\pi k-\pi/4)i/u$, то в каждой окрестности
${\cal U}_k$ этих точек
радиуса $C/u^2$, лежащей в $\Upsilon$
найдется и при том единственная точка спектра.
\end{lm}
\prf Проводится по той же схеме, что и доказательство леммы~\ref{thtalUpsk}. 
Отметим, что матрица перехода в этом случае строится на основании 
леммы~\ref{lm5.4}.

При фиксированном $\la\in\gamma$ линию Стокса, содержащую $\xi$ обозначим
через $l_1$. Дальнейшую нумерацию будем проводить против часовой стрелки.
Нумерацию канонических областей проведем аналогично условию 
леммы~\ref{lm5.4}. Для удобства обозначим $\eta$ --- оставшаяся (после $\xi$)
точка пары $a$, $b$. Соотношение на решения примет вид:
$$
\left(
\begin{array}{c}
U_1\\
V_1
\end{array}
\right)=
e^{-\frac{\pi}{6}i}
\left(
\begin{array}{cc}
0 & 1\\
1/\alpha_{12} & i\alpha_{23}
\end{array}
\right)
\left(
\begin{array}{c}
U_2\\
V_2
\end{array}
\right).
$$

Аналогично доказательству леммы~\ref{thtalUpsk}, получим соотношение:
$$
e^{-u(S_1(z_0,\xi)+S_2(z_0,\eta))}=\frac{1}{\alpha_{12}}
e^{u(S_1(z_0,\xi)+S_2(z_0,\eta))}(1+O(1/u))+
$$
$$
+i\alpha_{23}
e^{u(S_1(z_0,\xi)-S_2(z_0,\eta))}(1+O(1/u)),
$$
которое, в силу того, что $S_2(z_0,\eta)<0$ равносильно более простому:
$$
e^{2uS_1(z_0,\xi)}=-i(1+O(1/u)).
$$

Применяя к последнему соотношению теорему Руше, получим утверждение леммы.
\edlm

Имеет место следующее

\begin{pr}
Пусть кривая $\gamma$ характеризуется тем, что из пары точек $a$, $b$ найдется
одна (обозначим ее через $\xi$), являющаяся при всех $\la\in\gamma$ 
критической в смысте определения~\ref{krittochkaigraphstoksa}. Обозначим
$z_0(\la)$ --- точка поворота, из которой исходит линия Стокса содержащая
$\xi$. Тогда функция $\sa(\tilde\la_k,z_0(\tilde\la_k),\xi)$, рассматриваемая 
как функция аргумента $\la$ однолистна в некоторой малой окрестности $\gamma$.
\end{pr}

Рассмотрим функцию 
$$
K(\la)=e^{\pi i/4}\int\limits_a^b\sqrt{\zeta^2-\la}\,d\zeta.
$$

\begin{pr}
Функция $K$ является однолистной в об\-ласти $\Pi$.
\end{pr}

Доказательство приведено в работе~\cite{petrovskij}.

Для полноты картины опишем поведение спектра в окрестности кривой
$\tainf$.

\begin{lm}
\label{lm3ospectre}
Для любого интервала $I$ (конечного или бесконечного)
$I\subset \tainf$, $\la_2\not\in I$,
для любой окрестности $\Upsilon$ этого интервала, не содержащей некоторой окрестности
$\la_2$,
найдутся $u_0>0$ и число $C>0$ такие, что при всех $u>u_0$ окрестность
$\Upsilon$ будет содержать точки спектра.
Более того, если рассмотреть точки $\tilde\la_k$:
$K(\tilde\la_k)=i\pi k/u$, то в каждой
окрестности ${\cal U}_k$ этих точек
радиуса $C/u^2$, лежащей в
$\Upsilon$,
найдется и при том единственная точка спектра.
\end{lm}
\prf Уравнение на характеристический определитель, очевидно, эквивалентно
следующему:
$$
e^{2uK(\la)}=1+O(\frac{1}{u}),
$$
отбрасывая добавку, как и ранее, получим $2uK(\tilde\la_k)=2\pi k i$.
Пользуясь теоремой Руше, завершим доказательство.
\edlm

Отметим, что из следствия~\ref{cltalUpsk} и 
лемм~\ref{lm2ospectre},~\ref{lm3ospectre} следует следующая важная 

\begin{thrma}
\label{mnthch32}
Предельная спектральная функция (в смысле определения~\ref{df5.4}) для каждой из критических кривых предельного
спектрального графа представляется в виде
\begin{enumerate}
\item
В случае $\la\in\tapich$:
$$
N(\la)=\frac{1}{2\sqrt{\varepsilon}}e^{\frac{\pi}{4}i}\la.
$$
\item
В случае $\la\in\taa$:
$$
N(\la)=\frac{1}{\pi\sqrt{\varepsilon}}e^{-\frac{\pi}{4}i}\int\limits_{-\sqplus}^a
\sqrt{\zeta^2-\la}\,d\zeta.
$$
\item
В случае $\la\in\taam$:
$$
N(\la)=\frac{1}{\pi\sqrt{\varepsilon}}e^{-\frac{\pi}{4}i}\int\limits_{\sqplus}^a
\sqrt{\zeta^2-\la}\,d\zeta.
$$
\item
В случае $\la\in\tab$:
$$
N(\la)=\frac{1}{\pi\sqrt{\varepsilon}}e^{-\frac{\pi}{4}i}\int\limits_{\sqplus}^b
\sqrt{\zeta^2-\la}\,d\zeta.
$$
\item
В случае $\la\in\tainf$:
$$
N(\la)=\frac{1}{\pi\sqrt{\varepsilon}}e^{-\frac{\pi}{4}i}\int\limits_a^b
\sqrt{\zeta^2-\la}\,d\zeta.
$$
\end{enumerate}
\end{thrma}

Доказательство данного утверждения полностью повторяет доказательство 
аналогичной теоремы для случа симметричного профиля~\cite{petrovskij}.

Работа поддержана грантами РФФИ \No 010100691 и \No 001596100.
    
\end{document}